\documentclass[prl,amsmath,twocolumn,showpacs,superscriptaddress]{revtex4-1}
\usepackage{epsfig}
\usepackage{graphicx}
\begin{document}

\title{
Aging Wiener-Khinchin Theorem
 }

\author{N. Leibovich}
\author{E. Barkai}
\affiliation{Department of Physics, Institute of Nanotechnology and Advanced Materials, Bar Ilan University, Ramat-Gan
52900, Israel}

\pacs{05.40.-a,05.45.Tp}

\begin{abstract}
The Wiener-Khinchin theorem shows how the power spectrum of a stationary random signal $I(t)$ is related to its correlation function $\left\langle I(t)I(t+\tau)\right\rangle$.
We consider  non-stationary processes with the widely observed
aging correlation function $\langle I(t) I(t+\tau) \rangle \sim t^\gamma \phi_{\rm EN}(\tau/t)$ and relate it to the sample spectrum.
We formulate two aging Wiener-Khinchin theorems relating the power spectrum to the time and ensemble averaged
correlation functions, discussing briefly the advantages of each. 
When the scaling function $\phi_{\rm EN}(x)$ exhibits
a non-analytical behavior in the vicinity of its small argument we obtain aging $1/f$ type of spectrum. 
We demonstrate our results with three examples: blinking quantum
dots, single file diffusion and Brownian motion
in a logarithmic potential, showing that our approach is valid for a wide range of physical mechanisms. 
\end{abstract}

\maketitle

 Understanding how the strength of a signal is distributed in the frequency
domain, is central both in practical engineering problems and
in Physics. In many applications a random process $I(t)$ recorded in a 
time interval $(0,t_m)$ is analyzed with the sample spectrum
$S_{t_m}(\omega) = |\int_0 ^{t_m} I(t) \exp(- i \omega t) {\rm d} t|^2/t_m$,
which is investigated in the limit of a long measurement time $t_m$.
For stationary processes,  the fundamental
 Wiener-Khinchin theorem \cite{Kubo}
relates between the power spectrum density
and the correlation function $C(\tau) = \langle I(t) I(t+\tau) \rangle$
\begin{equation}
\lim_{t_m \to \infty} \langle S_{t_m}(\omega) \rangle = 2  \int_0 ^\infty C(\tau) \cos(\omega \tau) {\rm d}  \tau.
\label{eq01}
\end{equation}
However in recent years there is growing interest in the spectral
properties of non-stationary processes, where the theorem is not valid
\cite{Bouchaud,Margolin06,Eliazar,Aquino,Niemann,Rod,Bouchaud96,Silv,Schriefl}.
In general, there seems no point to discuss and classify
spectral properties of all possible non-stationary processes.
Luckily, a wide class of Physical systems and models exhibit a special
type of 
correlation
functions $\langle I(t) I(t+\tau) \rangle\sim
t^\gamma \phi_{{\rm EN}}(\tau/t)$ for an observable $I(t)$ and
the subscript ${\rm EN}$ denotes an ensemble average.  
Such correlation functions, describing what is  referred to as physical
aging,
 appear in a vast array of systems
and models ranging from glassy dynamics \cite{Bouchaud,Barrat,Dean,Bertin}, 
 blinking quantum dots \cite{Margolin04},
laser cooled
atoms \cite{DechantPRX13}, 
  motion of a tracer particle in a crowded environment \cite{Lizana,Leibovich}, 
elastic models of fluctuating interfaces \cite{Taloni}, 
deterministic noisy Kuramoto models \cite{Ionita},
granular gases \cite{Bordova}, and 
deterministic intermittency  \cite{AgingELI}, 
to name only a few examples.   
In some cases the scaling function exhibits a second scaling exponent,
$\langle I(t) I(t+\tau) \rangle\sim
t^\gamma \phi_{{\rm EN}}(\tau/t^\beta)$, or even
a logarithmic time dependence \cite{Bertin},
however here we will avoid this
zoo of exponents, and  attain classification of the spectrum 
for the case $\beta=1$.  

 A natural problem is to relate between the
sample spectrum of such processes and the underlying correlation function \cite{Margolin06}.
That such a relation actually exists is obvious from the basic
definition of the sample spectrum, see Eq. (\ref{eq02}) below.
However, here we find a few interesting insights. First, the
correlation function in its scaling form $\langle I(t) I(t + \tau) \rangle\sim t^\gamma \phi_{{\rm EN}}(\tau/t)$ is valid in Physical situations,
 in the limit of large $t$ and $\tau$. We here first formulate a theorem
for ideal processes, where the aging correlation function is valid for
all $\tau$ and $t$, and then in the second part of the Letter,
explore by comparison to realistic models the domain of validity
of the ideal models. As a rule of thumb the aging Wiener-Khinchin theorem
presented here for  ideal models works
well in the limit of low frequency. Further the limit of small
frequency and measurement time $t_m$ being large is not interchangeable
and should be taken with care. 
Secondly, the spectrum in these processes depends on time $t_m$, as already observed in \cite{Margolin06,Sadegh}. 
The non-stationarity 
also implies a third theme, namely that the ensemble average
correlation function is non identical to the time averaged correlation
function, in contrast with the usual Wiener-Khinchin scenario.
Thus we formulate two theorems, relating between
time and ensemble  average correlation functions and the sample spectrum.
The choice of theorem to be used in practice depends on the application.

In physics the power spectrum is not only a measure of the
strength of frequency modes in a system. Nyquist's fluctuation dissipation
theorem, for systems close to thermal equilibrium and hence stationary, 
states
that the ratio between the power spectrum and the imaginary part
of the response function, $\chi (\omega)$, is given by temperature, i.e.  $k_{B}T = \pi\omega S(\omega)/2\Im\left[\chi (\omega)\right]$ \cite{FDT.Kubo}. 
Similarly, effective temperatures are routinely defined by relating
measurements of power spectrum and response functions
of non-stationary processes \cite{LFC, Bellon, Crisanti}. Our goal here is to provide
the connection between the sample spectrum and the correlation
functions, without which the meaning of the
 effective temperature becomes
some what ambiguous. More practically, 
an experimentalist who uses the sample spectrum
to estimate the spectrum of a non-stationary process, might wish to
extract from it the time and/or the ensemble averaged correlation functions, 
and for that our work is valuable. 

{\em Aging Wiener-Khinchin theorem for time averaged correlation functions.}
For a general process, the autocorrelation function  $\langle I(t) I(t+\tau) \rangle$ is a function of its two variables, unlike stationary processes,
where the correlation function depends only on the time difference $\tau$.
Using the definition
of the sample spectrum we have
\begin{equation}
t_m \langle S_{t_m}(\omega) \rangle =  \int_0 ^{t_m} {\rm d} t_1 \int_0 ^{t_m} {\rm d} t_2 e^{ i \omega(t_2 - t_1)} \langle I(t_1) I (t_2) \rangle.
\label{eq02} 
\end{equation}
We identify in this equation the ensemble average correlation
function, but a formalism based on a time average will turn
out more connected to the original Wiener-Khinchin theorem,
as we proceed to show. A change of variable $\tau=t_2-t_1$ and
relabeling integration variables gives 
\begin{equation}
\langle S_{t_m} (\omega) \rangle = { 2 \over t_m} \int_0 ^{t_m} {\rm d} \tau (t_m - \tau) \langle C_{{\rm TA}} ( t_m, \tau)\rangle \cos\left( \omega \tau \right).
\label{eq03}
\end{equation}
Here the time averaged correlation function is defined as
\begin{equation}
C_{{\rm TA}} \left( t_m , \tau\right) = {1 \over t_m - \tau} \int_0 ^{t_m - \tau} {\rm d} t_1 I(t_1) I(t_1 + \tau). 
\label{eq04}
\end{equation}
We now insert in Eq. (\ref{eq03})
 an aging correlation function 
\begin{equation}
\langle C_{{\rm TA}} (t_m,\tau) \rangle = (t_m)^\gamma \varphi_{{\rm TA}} \left( \tau/ t_m \right),
\label{eq05}
\end{equation}
defining a new integration variable $0<\tilde{\tau} = \tau/t_m<1$ 
we find
\begin{equation}
\langle S_{t_m}(\omega) \rangle=2 (t_m)^{1 + \gamma} \int_0 ^1 {\rm d} \tilde{\tau} \left( 1 - \tilde{\tau} \right) \varphi_{{\rm TA}} \left( \tilde{\tau} \right) \cos\left(\omega t_m \tilde{\tau} \right).
\label{eq06}
\end{equation}
This formula relates between the time average correlation function
and the average of the sample spectrum, for ideal processes in the
sense that we have assumed that the scaling of the correlation function
holds for all times. 
 It shows
that the frequency $\omega$  times $t_m$ is the scaling variable of 
the power spectrum. 

{\em Aging Wiener-Khinchin formula for the ensemble averaged correlation
function.} 
We now relate the power spectrum with
the  ensemble
averaged correlation function which has a scaling form
\begin{equation}
\langle I(t + \tau) I(t) \rangle = t^\gamma \phi_{{\rm EN}}(\tau/t). 
\label{eq08}
\end{equation}
The two correlation functions are related with Eq. (\ref{eq04}), which upon averaging gives
\begin{equation}
 \varphi_{{\rm TA}}(x) = x^\gamma y(x) \int_{y(x)} ^\infty {\phi_{{\rm EA}}(z) \over z^{2+\gamma}}
{\rm d} z
\label{eq13a}
\end{equation}
 with $y(x)=x/(1-x)$.
Considering the case $\gamma=0$ we insert Eq. (\ref{eq13a}) in Eq. 
(\ref{eq06}) and find 
\begin{equation}
\left\langle S_{t_m}(\omega)\right\rangle=  
2t_m \int_0^1\phi_{{\rm EA}}\left(\frac{x} {1-x}\right){\tilde{\omega}x \sin(\tilde{\omega}x) +\cos(\tilde{\omega}x)-1\over \left(\tilde{\omega}x\right)^2} {\rm d} x
\label{eq10}
\end{equation}
with $\tilde{\omega} = \omega t_m$. 
For the  more general case $\gamma\ne 0$ we show
in  the supplementary material that
\begin{widetext}
\begin{equation}
\left\langle S_{t_m}(\omega)\right\rangle=
\frac{2(t_m)^{\gamma+1}}{2+\gamma}\int_0^1(1-x)^{\gamma}\phi_{{\rm EN}}\left(\frac{x}{1-x}\right) {_1F_2}\left(1+\frac{\gamma}{2};\frac{1}{2},2+\frac{\gamma}{2};-\left(\frac{\tilde{\omega}x}{2}\right)^2\right) {\rm d} x,
\label{eq09}
\end{equation}
\end{widetext}
where $_1F_2$ is a hypergeometric function and $\gamma>-2$.
This relation between the ensemble average correlation function and the sample averaged
spectrum is useful for theoretical investigations, when a microscopical
theory provides the ensemble average. 
Alternatively one may use the 
relation  Eq. (\ref{eq13a}) to compute the time average correlation function
from the ensemble average (if the latter is known) and then use the time averaged formalism Eq. (\ref{eq06}) which
is based on a simple cosine transform.  
The transformation
Eq. (\ref{eq09}) depends on $\gamma$, which in experimental situation
might be unknown (though it could be estimated from data), 
while Eq. (\ref{eq06}) does not, still both formalisms are 
clearly identical and
useful.

 {\em Relation with $1/f$ noise.}  We now consider a class of aging correlation
functions, with the additional characteristic behavior for small variable $\tau/t$
\begin{equation} 
\langle I(t) I(t + \tau) \rangle \sim  t^\gamma \left[ A_{{\rm EN}} - B_{{\rm EN}} \left( { \tau \over t}\right)^\nu \right].
\label{eq11}
\end{equation}
Here $A_{{\rm EN}},B_{{\rm EN}}>0,0<\nu<1,\gamma>-1$ and $\gamma-\nu>-1$. 
Physical examples
will soon follow. 
We use Eqs. (\ref{eq04},\ref{eq05},\ref{eq08},\ref{eq13a}) and by comparison
of coefficients of small argument expansion, we  show that
the time averaged correlation function has a similar expansion,
with $\langle C_{{\rm TA}} (t , \tau) \rangle \sim t^{\gamma} \left[ A_{{\rm TA}} - B_{{\rm TA}} (\tau/t)^\nu + \cdots \right]$
with $A_{{\rm TA}}= A_{{\rm EA}}/ (1 + \gamma)$ and $B_{{\rm TA}} = 
B_{{\rm EA}}/ (1 - \nu + \gamma)$. 
 We can then insert this
expansion in Eq. (\ref{eq06}), 
by integration by parts and using $\omega t_m = 2 \pi n\gg1$ where
$n$ is an integer
\begin{equation}
\langle S_{t_m}(\omega) \rangle \sim {2 \Gamma(1 + \nu) \sin \left( { \pi \nu \over 2} \right) B_{{\rm EA}} \over \left( \gamma - \nu +1 \right) (t_m)^{\nu - \gamma} \omega^{1 + \nu}}.
\label{eq12}
\end{equation}
%
We see that the non-analytical expansion of the
correlation function, in small argument, leads to a $1/f$ type
of noise, with an amplitude that depends on measurement time. 
Such an aging effect in the power spectrum was recently measured
for blinking quantum dots \cite{Sadegh}, so this shall be our first example. 

\begin{figure}\begin{center}
\includegraphics[width=0.50\textwidth]{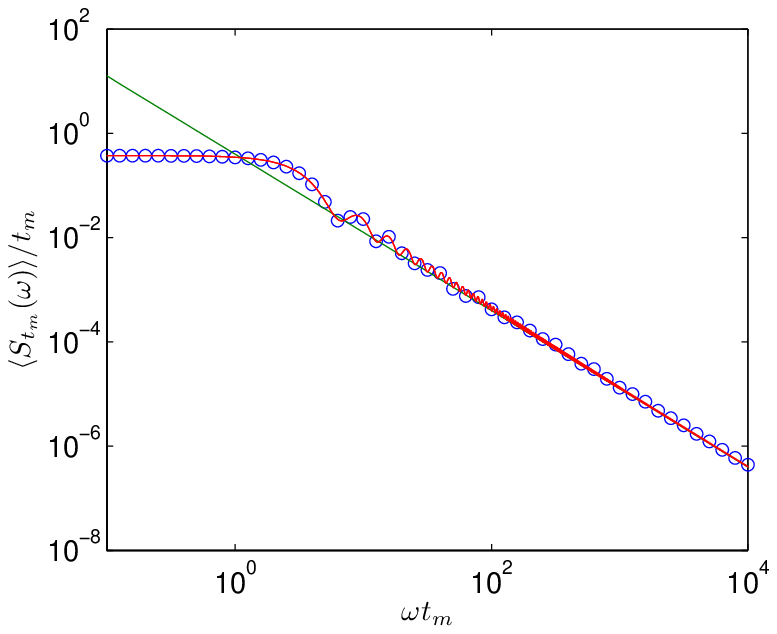}
\end{center}
\caption{(color online).
Power spectral density for the
 blinking quantum dot model with $\alpha=0.5$ with measurement time $t_m=10^5$. 
Theory Eq. (\ref{eq15}) (red line)
 perfectly matches finite time simulation (blue $\circ$) and asymptotically
the $1/f$ noise Eq. (\ref{eq16}) (green line).   
}
\label{fig1}
\end{figure}

{\em Blinking quantum dots and trap model.} As measurements show,
 blinking quantum dots, nano-wires and organic molecules exhibit episodes
of fluorescence intermittency, switching randomly between on and off states 
\cite{Kuno,Kraft,PhysTodayBarkai}.
 The on and off
waiting times are random  with a common  power law waiting time 
distribution 
 $\psi(\tau) \sim A \tau^{-(1 + \alpha)}$, a behavior valid
under certain conditions, like low temperature and weak external
laser field.
For this simple
renewal model, and when the average on and off times
diverge, namely  $0<\alpha<1$, we have
 $\gamma=0$ and  the correlation function, with intensity in the on state
taken to be $I_0$ and in the off state to be zero, is \cite{Margolin04}
\begin{equation}
\phi_{{\rm EN}} (x) =I_0^2 \left[{1 \over 2} - \frac{\sin(\pi\alpha)}{4\pi}  B\left( { x \over 1 + x}; 1- \alpha , \alpha\right)\right]
\label{eq13}
\end{equation}
where $x=\tau/t$ and $B(z;a,b)$ is the incomplete beta function.
Importantly, this type of correlation function describes not only blinking
dots, but also the trap model,
a well known model of glassy dynamics \cite{Dean}. 
The connection between the two systems
are the power law waiting times in micro-states of the system, though
for the trap model $\alpha=T/T_g$ where $T$ is temperature, while $0.5<\alpha<0.8$ in quantum dots experiments.
Eq. (\ref{eq13}) is valid only
in a scaling limit for large $\tau$ and $t$ when the microscopic
details of the model, e.g. the shape of the waiting time distribution $\psi(\tau)$
for short on and off blinking events, are irrelevant.
Thus the scaling solution is controlled only by the parameter $\alpha$. 
The time averaged correlation function is
obtained from Eqs. (\ref{eq13a},\ref{eq13})
\begin{eqnarray}
&&\phi_{{\rm TA}} (x) ={I_0^2 \over 4} + \label{eq14} \\ &&{I_0^2 \over 4}\frac{\sin(\pi\alpha)}{\pi} \left[
{ B\left(  1-x, \alpha ,1- \alpha \right)\over 1 -x} - {1 \over \alpha } \left( { x \over 1 - x}\right)^{1-\alpha} \right],
\nonumber 
\end{eqnarray}
which is  
clearly non-identical to the corresponding  ensemble averaged one.
We may now use either Eq. (\ref{eq06}) for the time average or
Eq. (\ref{eq10}) for the ensemble average to obtain $S_{t_m}(\omega)$.
Since  $\gamma=0$ we use Eq. (\ref{eq10}) and find
\begin{equation}
\langle S_{t_m}(\omega)\rangle/t_m =I_0^2 \left\{\frac{{\rm sinc}^2\left(\frac{\tilde{\omega}}{2}\right)}{4} + 
\frac{1}{2\tilde{\omega}}\Im\left[{M}(1-\alpha,2;\imath\tilde{\omega})\right]\right\},
\label{eq15}
\end{equation}
where $M(a,b;z)$ is the Kummer confluent hypergeometric function and $\Im\left[.\right]$ refers to its imaginary part. We note that the ${\rm sinc^2(\tilde{\omega}/2})$ term is the spectrum contribution from a constant $\left\langle \bar{I}^2\right\rangle$.    
As shown in Fig. \ref{fig1}
the spectrum Eq. \eqref{eq15} perfectly matches finite time 
 simulation of the process where we used $\psi(\tau) = \alpha \tau^{-(1 +\alpha)}$ for $\tau>1$, $\alpha=0.5$, $I_0=1$ and
an average over $10^3$ on-off blinking  processes was made. This
indicates that the scaling approach works well, even for reasonable
finite measurement time. The theory predicts nicely not only the generic
$1/f$ behavior but also the fine oscillations and the crossover to the
low frequency limit. As Fig. \ref{fig1} demonstrates when $\omega t_m$ is large, 
we get the $1/f$ noise result, which according to Eq. (\ref{eq12}) is
\begin{equation}
\langle S_{t_m}(\omega) \rangle  \sim { I_0^2\cos \alpha \pi /2 \over 2 \Gamma(1 + \alpha)} (t_m)^{\alpha -1} \omega^{\alpha-2}.
\label{eq16}
\end{equation}
  for $0<\alpha<1$.
In this model
$\nu= 1-\alpha$ as the small argument expansion of Eq. (\ref{eq13}) shows. 
The asymptotic Eq. (\ref{eq16}) 
 agrees with previous approaches \cite{Margolin06}, the latter missing the low frequency 
part of the spectrum, and the fine structure of the spectrum presented
in Fig. \ref{fig1},  since for those non trivial aspects of the theory
one needs the aging Wiener-Khinchin approach developed here. 
Finally, we have assumed that the start of the blinking process at $t=0$ (corresponding
to the switching  on of the laser field) is the moment in time where
we start recording the power spectrum. If one waits a time $t_w$ before
start of the measurement, the power spectrum will depend on the waiting
time $t_w$ since the process is non-stationary \cite{Bouchaud}. 

We note that a model with cutoffs on the aging behaviors was investigated
in \cite{Godec}, in this case the asymptotic behavior is normal, namely the Wiener-Khinchin theorem holds.  Indeed in experiments on blinking quantum
dots with a measurement time of $1800$ seconds the aging of the spectrum 
is still clearly visible 
\cite{Sadegh}, the latter measurement time is long in the sense that
blinking events are observed already on the $\mu$ Sec.  time scale. Hence cutoffs, while possibly important in some applications, are not relevant at least in this experiment.

\begin{figure}
\begin{center}
\includegraphics[width=0.50\textwidth]{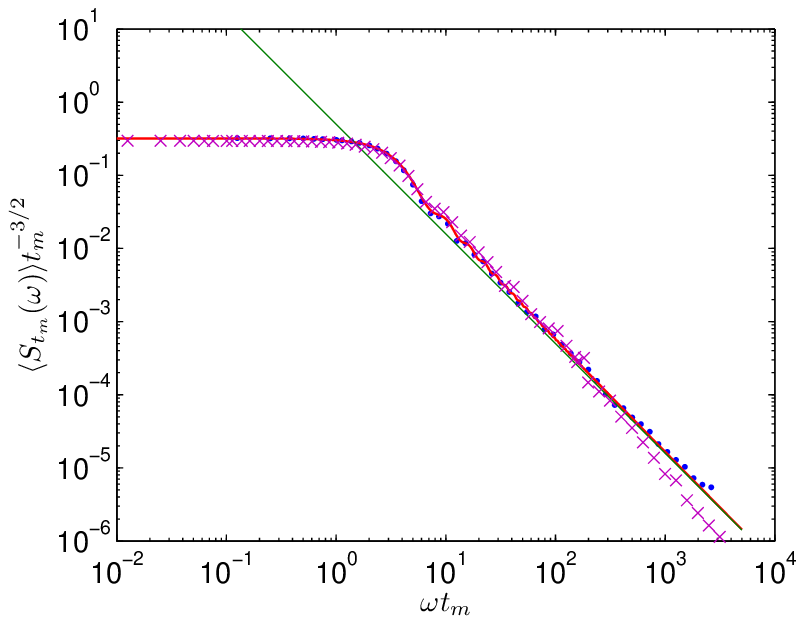}
\end{center}
                \caption{ (color online). The power spectrum of tagged particle motion $x(t)$ 
 with measurement
times  $t_m=10^3$ (blue $\bullet$) and $t_m=10^2$ (pink $\times$). 
For this single file model theory  Eq. (\ref{eq18}) (red line) 
and asymptotic $1/f$ approximation Eq. (\ref{eq20})
nicely match simulation results. 
In simulation we used $D=1/2$ and $a=1$. Finite time deviations for $t_m=10^2$ are observed at high frequency as discussed in text. 
}
\label{fig2}
\end{figure}

{\em Single file diffusion.} We consider a tagged Brownian particle
in an infinite unidimensional system, interacting with other
identical particles through hard core collisions \cite{Harris,Krapivsky,Hegde}. This well
known model of a particle in a crowded pore, is defined through
the free diffusion coefficient $D$ describing the motion of particles
between collision events and the averaged
distance between particles $a$. Initially at time
$t=0$ the particles  are uniformly distributed in space and
the tagged particle is on the origin. 
In this many body problem,
the motion of the tracer is sub-diffusive $\langle x^2(t) \rangle
\sim a \sqrt{D/\pi} \sqrt{t}$  since the other
particles are slowing down the tracer particles  via collisions \cite{Harris}.
Normal diffusion is found only at very short times $t<a^2/D$ when the tracer
particle has not yet collided with the other surrounding Brownian
particles. 
Our observable (so far called $I(t)$) is the position of the tracer 
particle
in space $x(t)$. The correlation function in the long time scaling limit
is \cite{Leibovich, Lizana}
\begin{equation}
\langle x(t) x(t + \tau) \rangle = a \sqrt{ { D \over \pi}} \sqrt{t} \left( \sqrt{ 1 + {\tau\over t}} + 1 - \sqrt{\tau\over t}\right).
\label{eq17}
\end{equation}
Such a correlation function describes also the coordinate of the Rouse
chain model, a simple though popular model of Polymer dynamics \cite{Rouse}.
Then by insertion and integration we find using Eqs. (\ref{eq06},\ref{eq13a})
\begin{eqnarray}
 (t_m)^{-3/2}\sqrt{\frac{\pi}{Da^2}}\left\langle S_{t_m}(\omega)\right\rangle&&=\frac{1}{\tilde{\omega}^{5/2}} \sqrt{\tilde{\omega}}(2+  \cos (\tilde{\omega}))   \label{eq18}\\  && -\frac{\sqrt{2\pi}}{2\tilde{\omega}^{5/2}}(1+2\cos(\tilde{\omega})){\cal{C}}(\sqrt{2\tilde{\omega}/\pi}) \nonumber \\ && +\frac{\sqrt{2\pi}}{\tilde{\omega}^{5/2}}{\cal{S}}(\sqrt{2\tilde{\omega}/\pi})(-\tilde{\omega}+\sin(\tilde{\omega})), \nonumber
\end{eqnarray}
where the Fresnel functions are defined as
${\cal{C}}(u)\equiv\int_0^u \cos(\pi t^2/2)dt$ and ${\cal{S}}(u)\equiv\int_0^u \sin(\pi t^2/2)dt$.
Generating  $10^3$ trajectories of single file motion, for a system with
$2001$ particles,  with the algorithm in \cite{Leibovich} we have found the sample spectrum
of the process $x(t)$.
As Fig. \ref{fig2} demonstrates theory and simulation for $\langle S_{t_m}(\omega)\rangle$
 perfectly match without fitting.  
From the correlation function Eq. (\ref{eq17})
we have $\gamma=\nu=1/2$ and hence according to
Eq. (\ref{eq12})
\begin{equation}
\langle S(\omega) \rangle \sim \sqrt{ {a^2 D \over 2} } \omega^{-3/2}
\label{eq20}
\end{equation}
for $\omega t_m \gg 1$. 
This equation  is the solid
line presented in Fig. \ref{fig2} 
 which is seen to match the exact theory already  for not too large
values of $\omega t_m$. 
 As in the previous example aging Wiener-Khinchin framework
 is useful
in the predictions of the deviations from the asymptotic result; as Fig. \ref{fig2}
clearly demonstrates some non-trivial wiggliness perfectly 
matching simulations. 

As mentioned in the introduction we have assumed a scaling form of the correlation function Eqs. (\ref{eq05},\ref{eq08}), which works in the limit of $t,\tau \rightarrow \infty$. 
Information on the correlation function for short times is needed to estimate the very high frequency limit of the spectrum. Hence the deviations at high frequencies in Fig. 2 are expected. As measurement time is increased the spectrum plotted as a function of $\omega t_m$ perfectly approaches the predictions of our theory (see also the following example
and Fig. 3). 

\begin{figure}
\begin{center}
\includegraphics[width=0.5\textwidth]{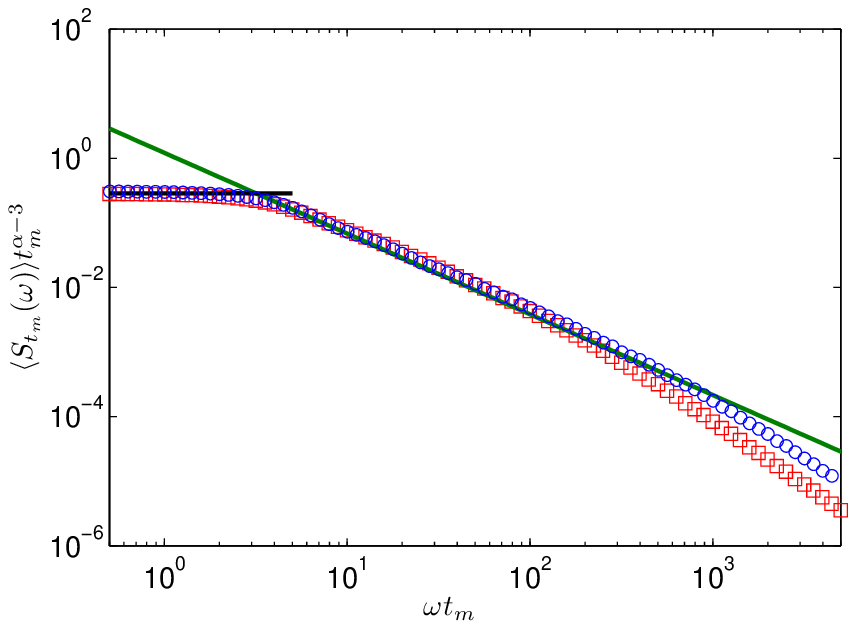}
\end{center}
\caption{(color online). 
Power spectrum of Brownian motion in a logarithmic potential (symbols) matches the asymptotic
theory predicting $1/f$ type of noise Eq. (\ref{eq22}) (green line). 
We use  $\alpha= 1.75,t_m=5\cdot10^2$ (red $\triangle$) and $t_m=2\cdot10^3$ (blue $\circ$) with $U_0=1$, $\bar{\gamma}=1$ and $m=1$. In addition we plot the theoretical zero frequency prediction $\langle S_{t_m}(0)\rangle$ (black line), see SM. The ensemble average was taken over $5000$ realizations. 
}
\label{fig3}
\end{figure}

{\em Diffusion in a logarithmic potential.} 
 While our previous examples
are  based on long tailed trapping times and many body interactions
which lead to a long term memory in the dynamics
 we will now briefly discuss a third mechanism
using  over damped  Langevin dynamics in a system which attains thermal
equilibrium. 
Consider the position $x(t)$, which is the observable $I(t)$, of
a particle with mass $m$  in a logarithmic potential $U(x) = U_0 \ln(1 + x^2)/2$ 
\begin{equation}
{{\rm d} x \over {\rm d} t} = - {1 \over m \bar{\gamma}} {\partial U\over \partial x} + \eta(t).
\label{eq21}
\end{equation}
Here the noise is white with mean equal zero satisfying the fluctuation dissipation theorem and $\bar{\gamma}$ is a friction constant. Under such conditions the
equilibrium probability density function is given by Boltzmann's law
$P_{{\rm eq}}(x) = \exp[ - U(x)/ k_b T]/Z$, where $Z$ is the
partition function and $T$ is the temperature. 
A key observation is that the potential is asymptotically 
weak in such a way that 
$P_{{\rm eq}}\sim x^{-U_0/k_b T}$ for large $x$  and for normalization to be
finite  we assume $U_0/k_b T> 1$. 
The system thus exhibits large fluctuations in its amplitude in the
sense that in equilibrium
 $\langle x^2 \rangle$ diverges in the regime 
$U_0 / k_b T < 3$. Of course for any finite measurement
 time the variance of $x(t)$,
starting on the origin, is increasing with time but finite.
Let $\alpha= U_0/(2 k_b T) + (1/2)$ and we focus on the case
$1<\alpha< 2$. The correlation function in this case was investigated
in \cite{Dechant2012}. 
We here study only the $1/f$ part of the spectrum,
demonstrating the versatility of the theory  using 
equation (\ref{eq12}), since unlike previous cases the correlation function is cumbersome. 
To find the $1/f$ noise we need to know, from the ensemble average correlation
function,  $\gamma,\nu$ and $B_{{\rm EN}}$,
while $A_{{\rm EN}}$ must be finite 
 (similar steps for other models
will be published elsewhere \cite{Notation}).
As detailed in the SM \cite{SM}  $\gamma=\nu=2-\alpha$ and
$B_{\rm EN}= \sqrt{\pi}(4D)^{2-\alpha}c_1/\left[Z\Gamma(\alpha)\Gamma(1+\alpha)\right] $
and $A_{\rm EN}= B_{\rm EN}\Gamma(1+\alpha)/\left[\sqrt{\pi}(2-\alpha) c_1\right]$
with $D=k_b T/ m\bar{\gamma}$ the diffusion constant according to the Einstein relation, hence by using Eq. \eqref{eq12} we obtain
\begin{equation}
\left\langle S(\omega)\right\rangle \sim    2 \Gamma(3 -\alpha) \sin\left({\alpha \pi\over 2}\right) B_{{\rm EN}}{1 \over \omega^{3-\alpha}} .
\label{eq22}
\end{equation}
The constant $c_1$ is given in terms of an integral of a special
function \cite{SM}. 
We have simulated the Langevin equation 
(\ref{eq21}) and obtained finite time estimates for the power spectrum,
which match the prediction Eq. (\ref{eq22}) as
shown in Fig. \ref{fig3} without fitting. Importantly,
the model of diffusion in logarithmic field is applicable in  many systems,
including diffusion of cold atoms in optical lattices \cite{Marksteiner}. 

{\em Summary and Discussion. } We have presented general relations between the sample spectrum and the time/ensemble averaged correlation function Eqs. (\ref{eq03},\ref{eq10}), respectively. Those relations work for Physical models in the limit $t_m\rightarrow\infty$ and $\omega\rightarrow 0$ while the product $\omega t_m$ is finite.
In experiment $t_m$ might be long, but it is always finite. Hence, the theorem will work in practice in the low frequency regime. Indeed, a close look at Fig. \ref{fig2}, \ref{fig3} shows finite time deviation at large frequencies, the aging spectrum is approached when $t_m$ is increased.
The fact that the scaled correlation function is observed in a great variety
of different systems, serves as evidence of the universality
of our main results, i.e. Eqs. (\ref{eq03},\ref{eq10}). 

\begin{acknowledgments}
This work was supported by the Israel science foundation.
\end{acknowledgments}

\end{document}